\documentclass[reprint, showpacs, superscriptaddress, aps, pre, twocolumn, 10pt]{revtex4-1}

\usepackage{graphicx} 
\usepackage{dcolumn}
\usepackage{bm}
\usepackage{amssymb}
\usepackage{amsmath}
\usepackage{wasysym}
\usepackage{color}
\usepackage{hyperref}
\usepackage{dcolumn}
\usepackage{float}
\usepackage{flushend}
\usepackage{multirow}
\usepackage{booktabs}

\hypersetup{
colorlinks,
citecolor=blue,
filecolor=blue,
linkcolor=blue,
urlcolor=blue}

\begin{document}
\title{A Shell Model for Buoyancy-Driven Turbulence}
\author{Abhishek Kumar}
\email{abhkr@iitk.ac.in}
\affiliation {Department of Physics, Indian Institute of Technology Kanpur, Kanpur, India 208016}
\author{Mahendra K. Verma}
\email{mkv@iitk.ac.in}
\affiliation {Department of Physics, Indian Institute of Technology Kanpur, Kanpur, India 208016}


\begin{abstract}
In this paper we present a unified shell model for stably stratified and convective turbulence.  Numerical simulation of this model for stably stratified flow shows Bolgiano-Obukhbov scaling in which the kinetic energy spectrum varies as $k^{-11/5}$.  We also observe a dual scaling ($k^{-11/5}$ and $k^{-5/3}$) for a limited range of parameters.  The shell model of convective turbulence yields Kolmogorov's spectrum.  These results are consistent with the energy flux and energy feed due to buoyancy, and are in good agreement with direct numerical simulations of Kumar {\em et al.} [Phys. Rev. E {\bf 90}, 023016 (2014)].
\end{abstract}

\maketitle
\section{Introduction}
\label{sec:intro}

Turbulence remains one of the unsolved problems of classical physics.  Turbulence generates strong nonlinear interactions among the large number of modes of the system, which makes theoretical analysis of such flows highly intractable.  Using Kolmogorov's theory of fluid turbulence, it can be shown that  the degrees of freedom of a turbulent flow with Reynolds number $\mathrm{Re}$ is $(\mathrm{Re})^{9/4}$~\cite{Frisch:book}.  Consequently,  a numerical simulation of a turbulent flow with a moderate Reynolds number of $\mathrm{Re} \approx 10^6$ requires $10^{27/2} \approx 31$ trillion grid points, which is impossible even on  the most sophisticated supercomputer of today.  

A low-dimensional model called {\em shell model of turbulence}~\cite{Ditlevsen:Book,Biferale:ARFM2003}  is reasonably successful in explaining certain features of turbulence, e.g., it reproduces the Kolmogorov's theory of fluid turbulence, as well as the experimentally observed intermittency corrections~\cite{Ditlevsen:Book,Biferale:ARFM2003}.   In a shell model, a single shell  represents all the modes of a logarithmically-binned shell, hence the number of modes in a shell model is much smaller than  $(\mathrm{Re})^{9/4}$.  Consequently, a large Reynolds number  can  be easily achieved in a shell model with 40 or more shells.

A large body of work exists on the shell model of fluid turbulence.  However, till date, there is no shell model for the stably stratified turbulence, and there is no convergence on the shell model for the convective turbulence.  Brandenburg~\cite{Brandenburg:PRL1992} and Mingshun and Shida~\cite{Mingshun:PRE1997}, have constructed shell models for convective turbulence, namely Rayleigh B\'{e}nard convection, but their results are divergent (to be discussed later; also see~\cite{Ching:PRE2008}).  In this paper, we introduce a shell model that describes both stably stratified and convective turbulence; we can go from one to the other with a change of sign in the density or temperature gradient.  Our shell model reproduces the numerical results of Kumar {\em et al.}~\cite{Kumar:PRE2014}, according to which stably stratified turbulence exhibits Bolgiano-Obukhov scaling~\cite{Bolgiano:JGR1959, Obukhov:DANS1959,Lohse:ARFM2010}, and convective turbulence shows Kolmogorov scaling. We also observe the dual spectrum predicted by Bolgiano~\cite{Bolgiano:JGR1959} and Obukhov~\cite{Obukhov:DANS1959}.

Buoyancy-induced turbulence~\cite{Lohse:ARFM2010}, often encountered in geophysics, astrophysics,  atmospheric and solar physics, engineering, come in two categories:   (a) Stably stratified flows in which a lighter fluid is above a heavier fluid.  These flows are stable because of their stable density stratification; (b)  Convective flows in which a heavier (or colder) fluid is above a lighter (or hotter) fluid.  Such flow configurations are unstable, hence, the heavier fluid  elements come down, and the lighter ones  go up.  The linear regimes of the above flow are easy to solve, and they yield gravity waves and convective instabilities respectively.  However, the turbulent aspects of such flows are active areas of research.

For stably stratified turbulence, Bolgiano~\cite{Bolgiano:JGR1959} and Obukhov~\cite{Obukhov:DANS1959} first proposed a phenomenology, according to which for $k<k_B$ ($k$ is wavenumber, and $k_B$ is Bolgiano wavenumber~\cite{Bolgiano:JGR1959}), the kinetic energy (KE) spectrum $E_u(k)$,  entropy spectrum $E_\theta(k)$,  KE energy flux $\Pi_u(k)$, and entropy flux $\Pi_{\theta}(k)$  are
\begin{eqnarray}
E_u(k) & =  & c_1 (\alpha^2 g ^2 \epsilon_\theta)^{2/5}k^{-11/5}, \label{eq:Eu} \\
E_\theta(k) & =  & c_2 (\alpha g)^{-2/5}\epsilon_\theta^{4/5} k^{-7/5}, \label{eq:Etheta} \\
\Pi_u(k) & = & c_3 (\alpha^2 g^2 \epsilon_\theta)^{3/5} k^{-4/5},  \label{eq:pi} \\
\Pi_\theta(k) & = &  \epsilon_\theta = \mathrm{constant}. \label{eq:pi_theta} 
\end{eqnarray}
Here  $\bf u$ and $\theta$ are the velocity and temperature fluctuations respectively,  $\alpha$ is  the thermal expansion coefficient, $g$ is the acceleration due to gravity, $\epsilon_{u}$ and $\epsilon_{\theta}$ are the  KE and entropy dissipation rates respectively, and $c_i$'s are constants.   The KE flux $\Pi_u(k)$ is forward, and it decreases with $k$ due to a conversion of  kinetic energy to  potential energy.  This decrease $\Pi_u(k)$ causes a steepening in the kinetic energy spectrum to $k^{-11/5}$, compared to  Kolmogorov's classical $k^{-5/3}$ spectrum.  Note that the stably stratified flows is also described in terms of density fluctuation $\rho'$, which leads to an equivalent description since $\rho' \propto -\theta$. In convective turbulence,  $\theta^2/2$ is referred to as the entropy, but in stably stratified turbulence, $\rho'^2/2$ is called the potential energy.

Bolgiano~\cite{Bolgiano:JGR1959} also showed that the buoyancy effects become somewhat insignificant in the wavenumber band $k_B<k<k_d$,  where $k_d$ is the dissipation wavenumber.  Therefore, $E_u(k), E_\theta(k) \sim k^{-5/3}$, $\Pi_u(k) = \epsilon_u$, and  $\Pi_\theta(k)=\epsilon_\theta$.  The aforementioned scaling, $k^{-11/5}$ for $k<k_B$, and  $k^{-5/3}$ for $k_B<k<k_d$, is referred to as Bolgiano-Obukhov (BO) phenomenology.    For convective turbulence, in particular for the idealised version called Rayleigh B\'{e}nard convection (RBC), Procaccia and Zeitak~\cite{Procaccia:PRL1989}, L'vov~\cite{Lvov:PRL1991},  L'vov and Falkovich~\cite{Lvov:PD1992}, and Rubinstein~\cite{Rubinstein:NASA1994} argued in favour of BO scaling.

The numerical and experimental findings however are inconclusive~\cite{Lohse:ARFM2010,Niemela:NATURE2000,Zhang:PRL2005,Mishra:PRE2010}.
In a recent numerical simulation, Kumar {\em et al.}~\cite{Kumar:PRE2014} analysed the above phenomenologies in a great detail.  They showed that stably stratified turbulence under strong buoyancy  exhibits $k^{-11/5}$ energy spectrum, as predicted by Bolgiano~\cite{Bolgiano:JGR1959} and Obukhov~\cite{Obukhov:DANS1959}.  They observed that $F(k) =  \mathrm{Re} (\langle u_k \theta^*_k\rangle) < 0$, thus corroborating the net conversion of  kinetic energy to potential energy.  For RBC, Kumar {\em et al.}~\cite{Kumar:PRE2014} showed that $F(k) =  \mathrm{Re} ( \langle u_k \theta^*_k\rangle) >0$, which is interpreted as a conversion of potential energy to kinetic energy, opposite to that in stably stratified turbulence.  As a result, for RBC,  the KE flux $\Pi_u(k)$ increases marginally with $k$, and KE exhibits an approximate Kolmogorov spectrum $k^{-5/3}$, contrary to the earlier predictions~\cite{Procaccia:PRL1989,Lvov:PRL1991,Lvov:PD1992,Rubinstein:NASA1994}.  The numerical results of Kumar {\em et al.}~\cite{Kumar:PRE2014} are consistent with those of Borue and Orszag~\cite{Borue:JSC1997}.

The powerlaw regimes in Kumar {\em et al.}'s~\cite{Kumar:PRE2014} simulations are somewhat narrow.  Also, they could not observe the dual spectrum of BO phenomenology, which may require much higher numerical resolution than $1024^3$.  In this paper, we present a unified shell model of stably stratified and convective turbulence that overcomes some of the aforementioned limitations of the numerical simulations.  For the unified shell model for the buoyancy driven flows,  we assume that the fluid is subjected to a mean temperature gradient, $d\bar{T}/dz$, which is positive for a stably stratified flow (cold below and hot above), and is negative for a convective flow (hot below and cold above).  Here $\bar{T}(z)$ is computed by averaging the temperature over the horizontal plane whose height is at $z$.

The outline of the paper is as follows. In Sec.~\ref{sec:model}, we introduce unified shell model, which can solve both stably stratified and convective turbulence; we discuss the construction of nonlinear terms, the required constraints, and the method for computing the  energy spectrum and flux.  Numerical results of the shell models for the stably stratified turbulence and the convective turbulence are discussed in Sec.~\ref{sec:SST} and ~\ref{sec:CT} respectively. We conclude in Sec.~\ref{sec:conclusions}.

\section{Shell Model for buoyancy-driven turbulence}
\label{sec:model}

Our shell model for the buoyancy-driven turbulence is
\begin{eqnarray}
\frac{du_{n}}{dt} & = & N_n[u,u] + \alpha g \theta_{n} - \nu k_{n}^{2}u_{n} + f_n, \label{eq:u_shell}\\
\frac{d\theta_{n}}{dt} & = & N_n[u,\theta] -  \frac{d \bar{T}}{dz} u_{n}  - \kappa k_{n}^{2}\theta_{n} \label{eq:T_shell},
\end{eqnarray}
where $u_n$ and $\theta_n$ are the shell variables for the velocity and temperature fluctuations respectively, $f_n$  represents the external force field,  $k_n = k_0 \lambda^n$ is the wavenumber of the $n$-th shell, and $\nu$ and $\kappa$ are the kinematic viscosity and thermal diffusivity, respectively, of the fluid. We choose $\lambda = (\sqrt{5}+1)/2$, the golden mean~\cite{Ditlevsen:Book}.

The nonlinear terms $N_n[u,u]$ and $N_n[u,\theta]$ are constructed keeping in mind the conservation of kinetic energy $\int d{\mathbf r} (u^2/2)$, kinetic helicity $\int d{\mathbf r} ( {\mathbf u}\cdot \mbox{{\boldmath $\omega$}})$, and entropy $\int d{\mathbf r} (\theta^2/2)$ in the absence of diffusive  and forcing terms.  For the shell model, the corresponding qualities are $\sum_n |u_n|^2/2$,  $\sum_n (-1)^nk_n|u_n|^2$,  and $\sum_n |\theta_n|^2/2$ respectively.  The nonlinear term $N_n[u,u] $ has been constructed earlier by invoking the conservation of kinetic energy and kinetic helicity as [see e.g,~\cite{Lvov:PRE1998}] 
\begin{eqnarray}
N_n[u,u] & = &-i(a_{1}k_{n}u_{n+1}^{*}u_{n+2}+a_{2}k_{n-1}u_{n-1}^{*}u_{n+1} \nonumber \\ 
& &-a_{3}k_{n-2}u_{n-1}u_{n-2}).
\end{eqnarray}
with constraints $a_1 + a_2 + a_3 = 0$ and $a_1 -\lambda  a_2 + \lambda^2 a_3 = 0$.  For our computation, we choose $a_1  = 1$, $a_{2} = \lambda -2$, and $a_{3} = 1 - \lambda$~\cite{Ditlevsen:Book}.  

For the construction of the nonlinear term $N_n[u,\theta]$, we use the fact that the nonlinear term of the temperature equation is a bilinear product of the temperature fluctuation and the velocity fluctuation.  Also,  the conservation of entropy yields a condition 
\begin{equation}
\mathrm{Re}\left( \sum \limits_{n} \theta_{n}^{*}N_n[u,\theta] \right)= 0.
\end{equation}
A combination of the above yields
\begin{eqnarray}
N_n[u,\theta] & = & -i[k_{n}(d_{1}u_{n+1}^{*}\theta_{n+2}+d_{3}\theta_{n+1}^{*}u_{n+2}) \nonumber \\
 &  & +k_{n-1}(d_{2}u_{n-1}^{*}\theta_{n+1}-d_{3}\theta_{n-1}^{*}u_{n+1}) \nonumber\\
 &  & -k_{n-2}(-d_{1}u_{n-1}\theta_{n-2}-d_{2}\theta_{n-1}u_{n-2})]
 \label{eq:nlin_T} 
\end{eqnarray}
with arbitrary $d_1,d_2$, and $d_3$.  For our shell model, we choose $d_{1}= 1$, $d_{2} = \lambda -2$, and $d_{3} = 1 - \lambda$.   For consistency, we choose the boundary conditions  $u_{-1}=u_0=\theta_{-1}=\theta_0 = 0$ and $u_{{N}+1}=u_{{N}+2}=\theta_{{N}+1}=\theta_{{N}+2} = 0$, where ${N}$ is the total number of shells.   Also note that we use Sabra model~\cite{Lvov:PRE1998} that yields less fluctuations for the spectrum compared to the GOY model~\cite{Gledzer:SPD1973,Lvov:PRE1998,Biferale:ARFM2003}.

The second term in the RHS of Eq.~(\ref{eq:u_shell}), $ \alpha g \theta_n$, is the buoyancy term, while $-(d\bar{T}/dz) $ is the temperature stratification (or equivalently density stratification) term.  Clearly, $d\bar{T}/dz > 0$ for a stably stratified flow, and  $d\bar{T}/dz < 0$ for the convective turbulence.  

The shell model for RBC does not require forcing to maintain  a steady state. However, the shell model for the stably stratified turbulence requires a forcing for the same; we force a set of small wavenumber shells (large length-scale modes) randomly so as to feed a constant energy supply rate $\varepsilon$ to the system.  We assume that the forcing shells receive equal amount of energy.  If $n_f$ shells are forced, then the above conditions yield the force at the $n$-th shell as
\begin{equation}
f_n = \sqrt{\frac{\varepsilon }{n_f  \Delta t}} e^{i\phi_n},
\end{equation}
where $\phi_n$  is the random phase of the $n$-th shell chosen from the uniform distribution in $[0,2\pi]$.  In our simulation we force the shells $n=3$ and 4, hence $n_f=2$.

  It is convenient to work with the nondimensionalized equations, which is achieved by using box height or the characteristic length $d$ as the length scale, $\sqrt{\alpha g | d\bar{T}/{dz} | d^2}$ as the velocity scale, and $|d\bar{T}/{dz}| d$ as the temperature scale. Therefore, $u_n = u^{\prime}_n \sqrt{\alpha g | d\bar{T}/{dz} | d^2}$, $\theta_n = \theta^{\prime}_n  | d\bar{T}/{dz} | d$,  $k_n = k^{\prime}_n / d$, and $t = t^{\prime} (d/\sqrt{\alpha g | d\bar{T}/{dz} | d^2})$. In terms of nondimensonalized variables, the equations are 
\begin{eqnarray}
\frac{du^{\prime}_{n}}{dt^{\prime}} & = & N^{\prime}_n[u^{\prime},u^{\prime}] + \theta^{\prime}_{n} - \sqrt{ \frac{\mathrm{Pr}}{\mathrm{Ra}}} {k^{\prime 2}_{n}}u^{\prime}_{n} + f^{\prime}_n, \label{eq:non_u_shell}\\
\frac{d\theta^{\prime}_{n}}{dt^{\prime}} & = & N^{\prime}_n[u^{\prime},\theta^{\prime}] -  S u^{\prime}_{n}  -   \frac{1}{\sqrt{\mathrm{Ra}\mathrm{Pr}}}k^{\prime 2}_{n}\theta^{\prime}_{n} \label{eq:non_T_shell},
\end{eqnarray}
where $S = 1$  for positive $ | d\bar{T}/{dz} |$, and $S = -1$  for negative $ | d\bar{T}/{dz} |$.

We remark that ours is the first shell model for the stably stratified turbulence.  For RBC, Brandenburg~\cite{Brandenburg:PRL1992}, and Mingshun and Shida~\cite{Mingshun:PRE1997} had constructed shell models.  Our shell model differs quite significantly from that of Brandenburg.  The shell model ``2" of Mingshun and Shida~\cite{Mingshun:PRE1997} is applicable to neutral stratification, and it is a subset of our shell model.  The shell model of Ching and Cheng~\cite{Ching:PRE2008} is same as that of Brandenburg~\cite{Brandenburg:PRL1992}.

The important parameters for the buoyancy-driven turbulence are:  the Prandtl number $\mathrm{Pr} = \nu/\kappa$, the Reynolds number is $\mathrm{Re} =u_\mathrm{rms} d/\nu$, and 
\begin{eqnarray}
\text{Brunt V\"{a}is\"{a}l\"{a} freq.}~N_f & = &  \sqrt{ \frac{g}{\rho_0}  \left| \frac{d \bar{\rho}}{dz} \right|} \nonumber \\ 
& = &\sqrt{\alpha g \left| \frac{d\bar{T}}{dz} \right|} \\
\mathrm{Rayleigh~number}~\mathrm{Ra} & = &\frac{d^4 \alpha g}{\nu \kappa}\left| \frac{d\bar{T}}{dz} \right|   \\
\mathrm{Froude~number}~\mathrm{Fr} & = & \frac{u_\mathrm{rms}}{d N_f}    \\
 \mathrm{Richardson~number}~\mathrm{Ri} & = & \frac{\alpha g d^2 }{u_{\rm rms}^2} \left| \frac{d\bar{T}}{dz} \right|  
\end{eqnarray}
where $u_\mathrm{rms}$  is the rms velocity of flow, and $d$ is the characteristic length scale. Note that the Brunt V\"{a}is\"{a}l\"{a} frequency is the frequency of the gravity wave,  the Froude number is the ratio of the characteristic fluid velocity and the gravitational wave velocity, and the Richardson number  is the ratio of the buoyancy and the nonlinearity $(\bf u \cdot \nabla) \bf u$. For convenience, the primes from the variables are dropped in our subsequent discussion.

Note that for RBC, the critical Rayleigh number $\mathrm{Ra}_{c} = 1$, after which the flow becomes unstable.  Due to the lower critical Rayleigh number in the shell model,  turbulence appears at a lower Rayleigh number compared to that observed in direct numerical simulations.  We compute energy spectrum  [$E_u(k)$] and the entropy spectrum [$E_{\theta}(k)$], defined as 
\begin{eqnarray}
E_u(k) & = & \frac{|u_k|^2}{k} ,\\
E_{\theta}(k) & = & \frac{| \theta_k|^2}{k},
\end{eqnarray}
using the steady state data.

We also compute the energy and entropy fluxes.  In fluid turbulence, the energy flux $\Pi(k)$ is defined~\cite{Verma:PR2004} as the rate of energy transfer from modes inside a sphere of radius $k$ to the modes outside the sphere. For the shell model, the energy flux $\Pi(k)$ is the rate of energy transfer from the shells within the sphere of radius $k$, i.e. $m\in[0,k]$,  to the shells outside the sphere, i.e. $n\in(k,N]$~\cite{Ditlevsen:Book}:
\begin{eqnarray}
\Pi_{u}(k) & = & \sum_{n>k}\sum_{m\le k}\sum_{p} -k_p \mathrm{Im}(u_p u_m u_n^{*}),
 \label{eq:Pi_KE}
\end{eqnarray}
Similarly, the entropy flux $\Pi_{\theta}(k)$ is defined as the rate of entropy transfer from the shells within the sphere of radius $k$ to the shells outside the sphere, i.e.,
\begin{eqnarray}
\Pi_{\theta}(k) & = & \sum_{n>k}\sum_{m\le k}\sum_{p} -k_p \mathrm{Im}(u_p \theta_m \theta_n^{*}).
 \label{eq:Pi_T}
\end{eqnarray}

We simulate the aforementioned shell model [Eqs.~(\ref{eq:non_u_shell}, \ref{eq:non_T_shell})] for stably-stratified and convective turbulence and compute the above spectra and fluxes.  We take $36$ shells for the stably stratified turbulence simulations SST1 and SST2, and $76$ shells for SST3 and the convective turbulence simulation (CT). For time stepping, we use the fourth-order Runge-Kutta (RK4) method.  For stably stratified turbulence, we apply random force on shells $n=3$ and $n=4$. The parameters of the simulations are listed in Table~\ref{table:simulation_details}.

\setlength{\tabcolsep}{11pt}
\begin{table*}[htbp]
\begin{center}
\caption{Parameters of our simulations: Flow type [stably stratified turbulence (SST) or convective turbulence (CT)], number of shells ${N}$, Rayleigh number $\mathrm{Ra}$, energy supply rate $\varepsilon$, Reynolds number $\mathrm{Re}$, Richardson number $\mathrm{Ri}$, Froude number $\mathrm{Fr}$, kinetic energy dissipation rate $\epsilon_u$, entropy dissipation rate $\epsilon_{\theta}$,  and Bolgiano wave number $k_B$. We choose $\mathrm{Pr}=1$ for all our runs.}
\begin{tabular}{c  c c c c c c c c c  }
\hline \hline \\[0.3 pt]
Flow Type & ${N}$  & $\mathrm{Ra}$ & $\varepsilon$ & $\mathrm{Re}$ & $\mathrm{Ri}$ & $\mathrm{Fr}$  &  $\epsilon_u$ & $\epsilon_{\theta}$ & $k_B$ \\[2 mm]
\hline \\[0.5 pt]
SST1 & $36$ 	& $10^{5}$ 	& $50$ 	& $1.0 \times 10^3$ 					& $0.10$ 					& $3.2$ 					&$3.9$& $3.1$ &$53$ \\[2 mm]
{SST2} & $36$ 	& $10^{10}$	& $10$ & $2.0 \times 10^5$ 	& $0.25$ 					& $2.0$		&			$0.6$ & $0.8$ & $18$		\\[2 mm]
SST3 & $76$ 	& $10^{5}$ 	& $10^{10}$ 	& $7.9 \times 10^5$ 	& $1.6 \times 10^{-7}$ 	& $2.5 \times 10^3$ 	&$1.4  \times 10^9$	& $7.6 \times 10^3$ & $<1$\\[2mm]
CT 	& $76$	& $10^{12}$ & NA 	& $8.7 \times 10^6$ 	& $0.01$ 					& NA 						&$62$					& $60$ 					& NA\\
\hline \hline
\end{tabular}
\label{table:simulation_details}
\end{center}
\end{table*}

We compute the spectra and fluxes of KE and entropy, and  average over many snapshots ($ \sim 10^{8}$) of the steady-state flow (of a single run); these values are further averaged  over $100$ simulations with independent random initial conditions~\cite{SankarRay:2008bc,Chakraborty:2010ij}.  The error bars reported in the paper for the spectral exponents and fluxes are  the standard deviations of the aforementioned 100 independent data sets~\cite{SankarRay:2008bc,Chakraborty:2010ij}.  This is the statistical error of our data. The spectral exponents of the energy and entropy spectra along with the errors  for the sets of parameters are summarized in Table~\ref{table:spectral_exponent}.  The data also has some systematic error, for example the dip in energy spectrum near $k\approx 7$ (the fifth shell).  The origin of the dip is not  understood clearly at present, and it will be presented in future.

\section{Energy and Fluxes of Stably Stratified Turbulence}
\label{sec:SST}

\begin{figure}[htbp]
\begin{center}
\includegraphics[scale = 1]{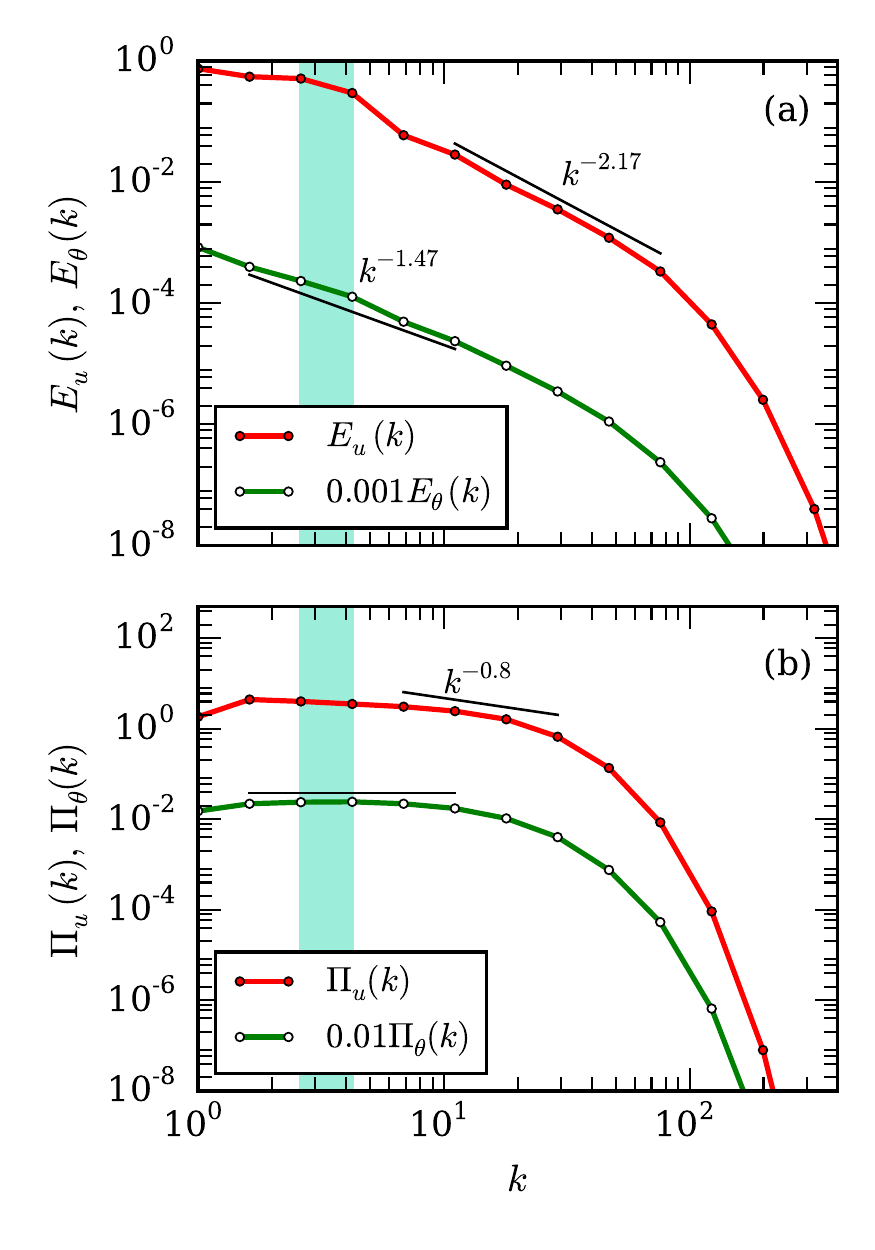}
\end{center}
\setlength{\abovecaptionskip}{0pt}
\caption{(Color online) For stably stratified simulation with $\mathrm{Pr} = 1$, $\mathrm{Ra} = 10^{5}$, and $\mathrm{Ri} = 0.10$: (a) plots of  KE and entropy spectra; (b) plots of KE flux $\Pi_u(k)$ and entropy flux $\Pi_{\theta}(k)$. The green shaded region shows the forcing range.}
\label{fig:spectrum_flux_ra_1e5}
\end{figure}

To test the validity of BO scaling in  the stably stratified turbulence, we simulate the shell model for the three sets of parameters, SST1, SST2, and SST3, which are listed in Table~\ref{table:simulation_details}. In Fig.~\ref{fig:spectrum_flux_ra_1e5}{\color{blue}(a)} we plot the KE spectrum $E_u(k)$ and entropy spectrum $ E_\theta(k)$ for $\mathrm{Ra} = 10^{5}$ and $\mathrm{Ri} = 0.10$. Green shadow regions in all the figures of the paper are the forcing bands. The figure indicates that $E_u(k) \sim k^{-2.17}$ and $E_\theta(k)  \sim k^{-1.47}$ for more than a decade, a result consistent with the BO scaling.

\begin{figure}[htbp]
\begin{center}
\includegraphics[scale = 1]{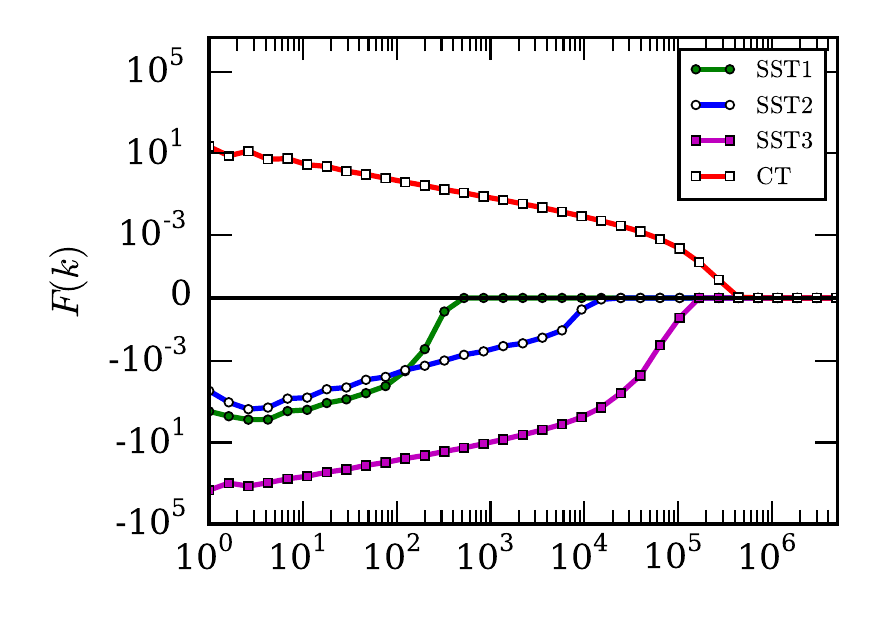}
\end{center}
\setlength{\abovecaptionskip}{0pt}
\caption{(Color online) (a) Plots of $F(k)$ for stably stratified turbulence SST1, SST2, SST3, and for convective turbulence (CT); $F(k)<0$ for SST's, but $F(k)>0$ for CT. }
\label{fig:force_spectrum}
\end{figure}

We also compute the KE and potential energy fluxes, which are plotted in Fig.~\ref{fig:spectrum_flux_ra_1e5}{\color{blue}(b)}. In the inertial range,  the entropy flux $\Pi_\theta(k)$ is constant, and the KE flux $\Pi_u(k)$ decreases with $k$, but somewhat different from $k^{-4/5}$.  These results are in general agreement with the BO scaling for the stably stratified turbulence.   We also compute energy supply rate by buoyancy, $F(k) = \mathrm{Re} ( \langle u_k \theta^*_k\rangle)$, which is negative, as shown in Fig.~\ref{fig:force_spectrum}.  Thus, we show a conversion of kinetic energy to  potential energy by buoyancy, a result consistent with that of Kumar  {\em et al.}~\cite{Kumar:PRE2014}.

\begin{figure}[htbp]
\begin{center}
\includegraphics[scale = 1]{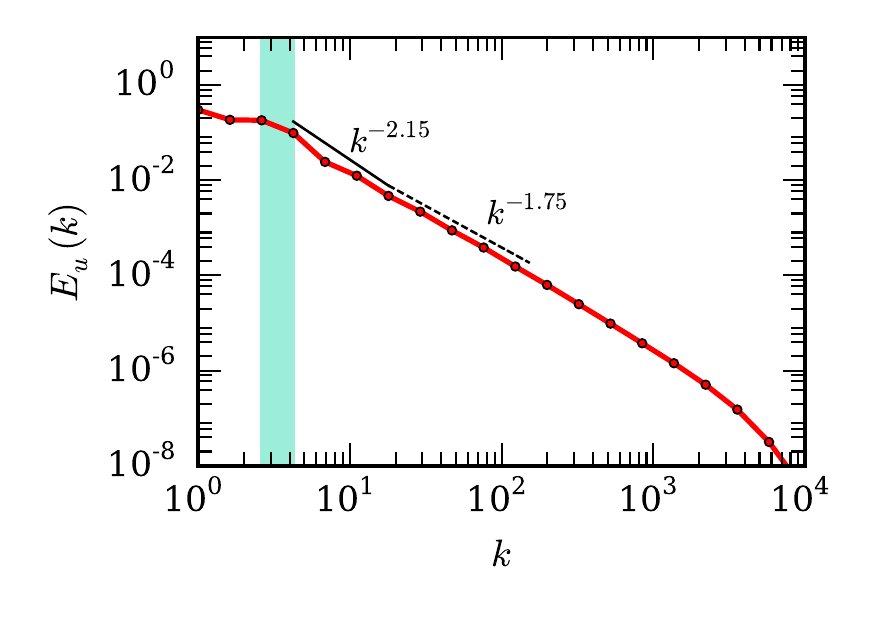}
\end{center}
\setlength{\abovecaptionskip}{0pt}
\caption{(Color online) For stably stratified simulation with $\mathrm{Pr} = 1$, $\mathrm{Ra} = 10^{10}$, and $\mathrm{Ri} = 0.25$, plot of KE spectrum. The wavenumber range $4 < k < 18$ exhibit $E(k)\sim k^{-11/5}$,  and $18 < k <100$ exhibit $E(k)$ close to the Kolmogorov spectrum.}
\label{fig:spectrum_ke_ra_1e10}
\end{figure}

\setlength{\tabcolsep}{11pt}
\begin{table}[htbp]
\begin{center}
\caption{Spectral exponents of our simulations for the runs SST1, SST2, SST3, and CT listed in Table 1:  KE spectrum $E_u(k) \sim k^{-p}$ and the entropy spectrum $E_\theta(k) \sim k^{-q}$. The two exponents $p,q$'s for SST2 are for the dual $BO$ scaling.}
\begin{tabular}{c   c c  }
\hline \hline \\[0.3 pt]
Flow Type & $p$ &$q$\\[2 mm]
\hline \\[0.5 pt]
SST1  &$-2.1677 \pm 0.0004$ &$-1.4667 \pm 0.0008$\\[2 mm]
\multirow{2}{*}{SST2} 	& $-2.147 \pm 0.004$;  & $-1.117\pm 0.009$; \\&$-1.746 \pm 0.001$ & $-1.6826 \pm 0.0009$\\[2 mm]
SST3 &$-1.7513 \pm 0.0006$ &$-1.6868 \pm 0.0007$\\[2mm]
CT 	& $-1.703 \pm 0.003$ &$-1.711 \pm 0.003$\\
\hline \hline
\end{tabular}
\label{table:spectral_exponent}
\end{center}
\end{table}

For the above case, the Bolgiano wavenumber $k_B \approx 53$, which lies in the dissipation range, thus making the $k^{-5/3}$ KE spectrum inaccessible. The Bolgiano wavenumber $k_B$ is calculated by comparing Eq.~(\ref{eq:Eu}) and the Kolmogorov's KE energy spectrum~\cite{Lohse:ARFM2010}. To obtain the dual spectrum predicted in BO scaling, we increase the Rayleigh number to $10^{10}$, which yields $\mathrm{Re} = 2.0 \times 10^5$, $\mathrm{Ri} = 0.25$, and $\mathrm{Fr}=2.0$ (SST2 of Table~\ref{table:simulation_details}).  For these parameters, we observe an approximate dual spectrum, as shown in   Fig.~\ref{fig:spectrum_ke_ra_1e10}.  The KE spectrum can be approximately described by $E_u(k) \sim k^{-11/5}$ for $4 < k < 18$, and $E_u(k) \sim k^{-5/3}$  and $18 < k <100$ respectively, with Bolgiano wavenumber $k_B = 18$.   Thus our simulations confirm the presence of dual scaling in stably stratified turbulence, as predicated by Bolgiano~\cite{Bolgiano:JGR1959} and Obukhov~\cite{Obukhov:DANS1959}. We observe that a further increase of $\mathrm{Ra}$ shrinks the $11/5$ regime and makes it invisible.

\begin{figure}[htbp]
\begin{center}
\includegraphics[scale = 1]{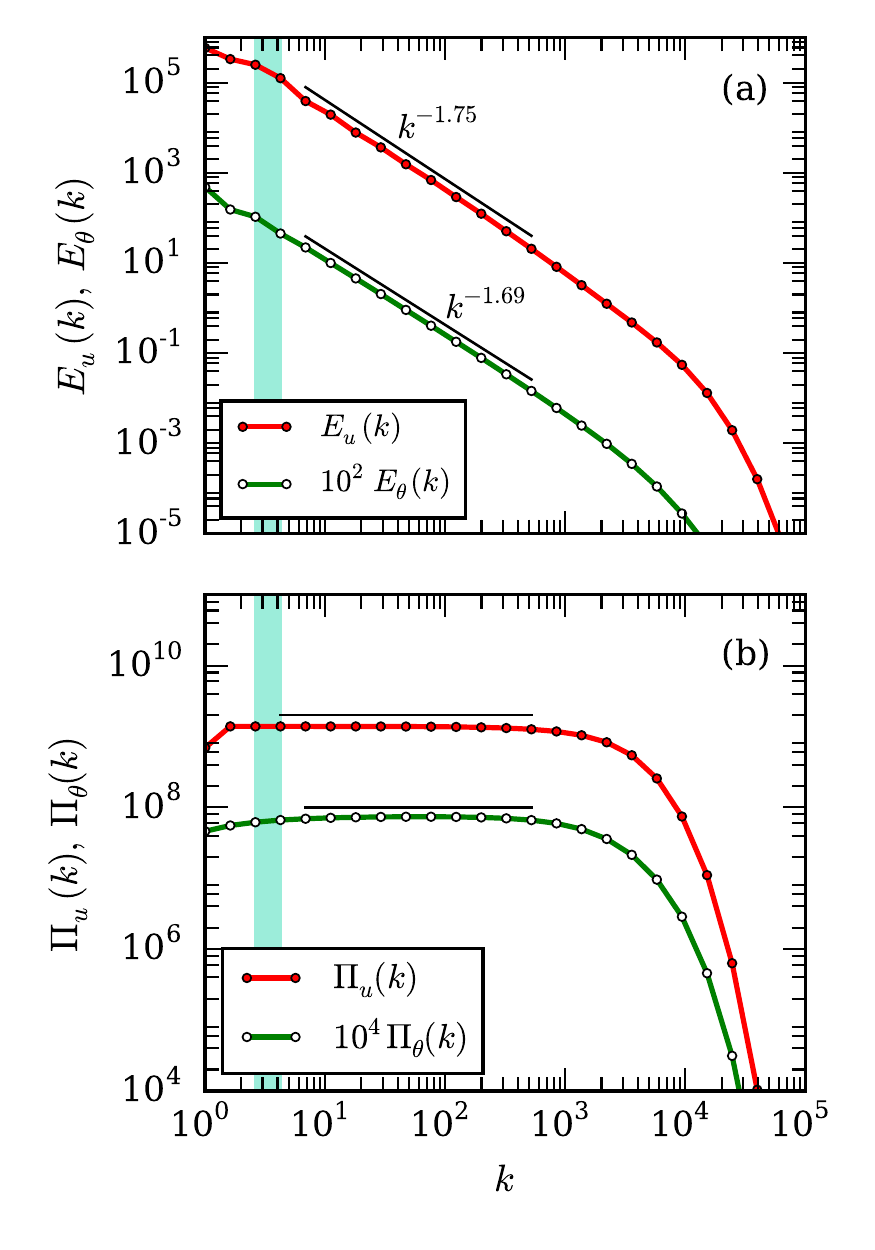}
\end{center}
\setlength{\abovecaptionskip}{0pt}
\caption{(Color online) For stably stratified simulation with $\mathrm{Pr} = 1$, $\mathrm{Ra} = 10^{5}$, and $\mathrm{Ri} = 1.6 \times 10^{-7}$: the (a) plots of  KE and entropy spectra; and (b) plots of KE flux $\Pi_u(k)$ and entropy flux $\Pi_{\theta}(k)$ exhibit Kolmogorov's spectrum since Ri or buoyancy is small.   }
\label{fig:spectrum_flux_ra_1e5_low_ri}
\end{figure}

For the parameters of SST3, the nonlinearity is stronger than the buoyancy term, which is evident from the fact the Richardson number $\mathrm{Ri} \ll 1$.   Hence, we observe Kolmogorov scaling, i.e. $E_u(k) \sim k^{-5/3}$ and $E_\theta(k) \sim k^{-5/3}$ for these parameters shown in Fig.~\ref{fig:spectrum_flux_ra_1e5_low_ri}{\color{blue}(a)}. The fluxes of KE and potential energy are constant in $k$,  as shown in Fig.~\ref{fig:spectrum_flux_ra_1e5_low_ri}{\color{blue}(b)}.

In the next section we will discuss the results of convective turbulence.

\section{Energy and Fluxes of Convective Turbulence}
\label{sec:CT}
In convective turbulence, buoyancy feeds energy to the kinetic energy, hence the KE flux increases marginally at lower wavenumbers~\cite{Kumar:PRE2014}.  In the intermediate range of wavenumbers, where the dissipation rate $D(k)= \sum\limits_{k} 2 \nu k^2 |u_k|^2/2$ approximately balances the energy supplied by buoyancy $F(k)$, we expect  Kolmogorov's spectrum for the velocity field~\cite{Kumar:PRE2014}.  We performed a shell model calculation to verify the above conjecture using the parameters  $\mathrm{Pr}=1$ and $\mathrm{Ra}=10^{12}$ (CT of Table~\ref{table:simulation_details}).  Note that no external forcing is required to obtain a steady state in convective turbulence. 

\begin{figure}[htbp]
\begin{center}
\includegraphics[scale = 1]{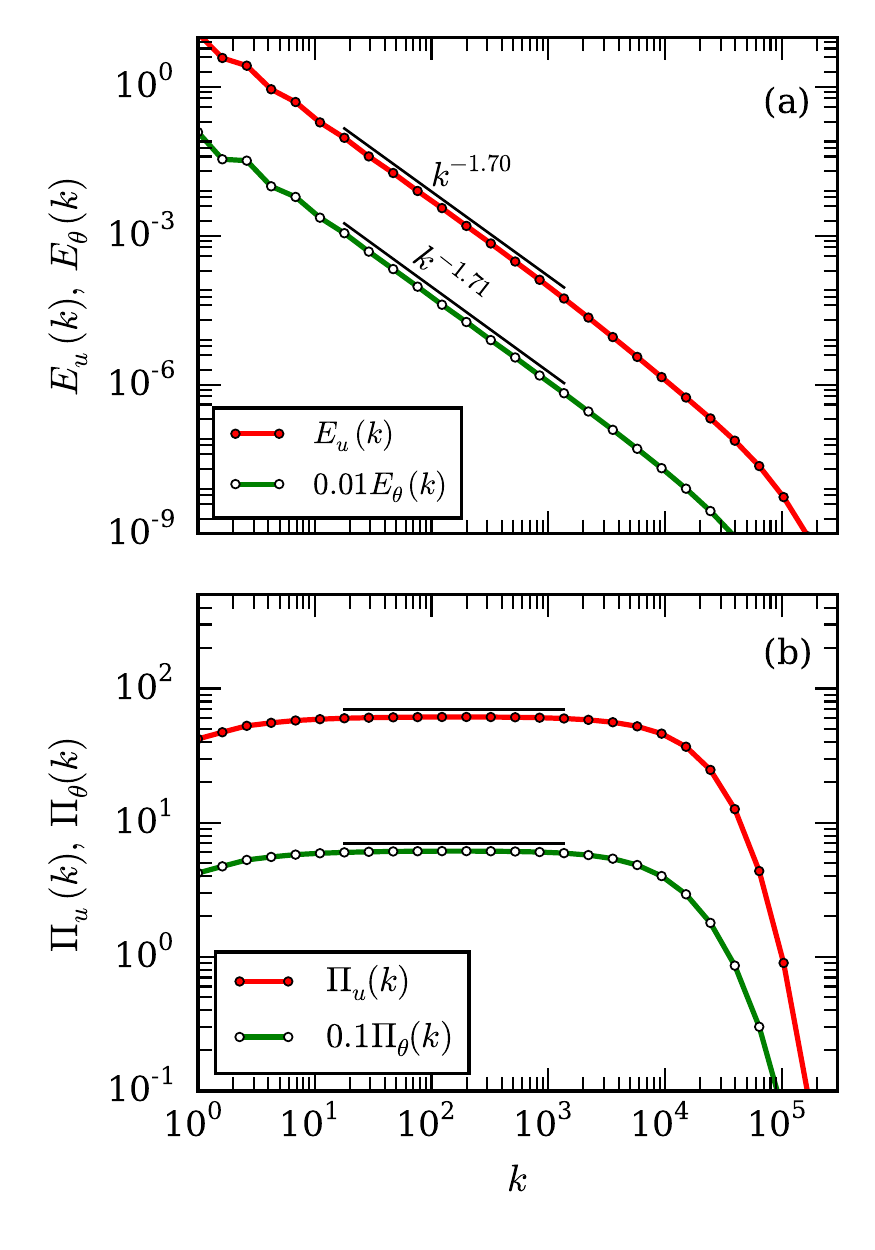}
\end{center}
\setlength{\abovecaptionskip}{0pt}
\caption{(Color online) For convective turbulence simulation with $\mathrm{Pr} = 1$ and $\mathrm{Ra} = 10^{12}$, the (a) plots of  KE and entropy spectra; (b) plots of KE flux $\Pi_u(k)$ and entropy flux $\Pi_{\theta}(k)$  exhibit Kolmogorov's spectrum. }
\label{fig:spectra_flux_RB}
\end{figure}

\begin{figure}[htbp]
\begin{center}
\includegraphics[scale = 1]{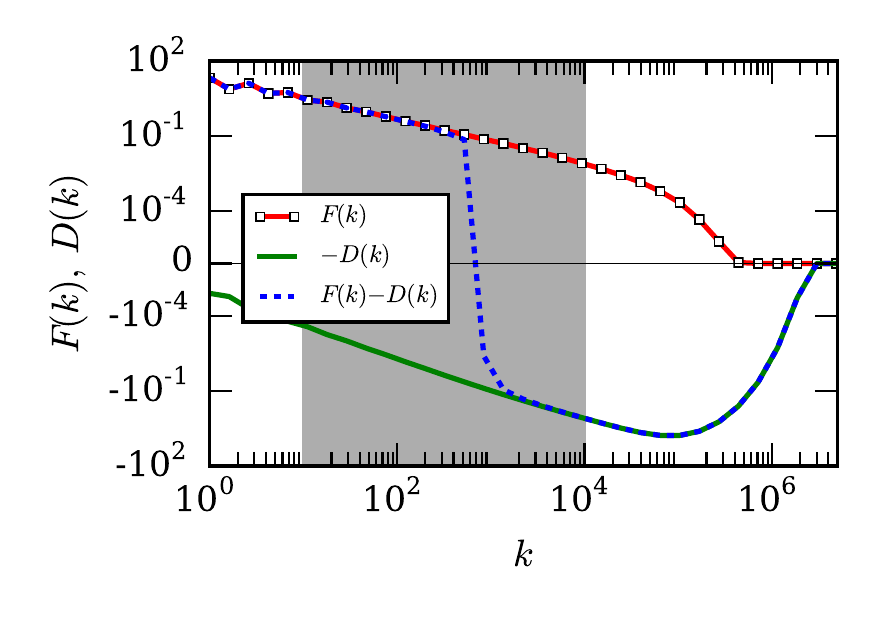}
\end{center}
\setlength{\abovecaptionskip}{0pt}
\caption{(Color online) For RBC run (CT), plots of  $F(k)$, $-D(k)$, and $F(k)-D(k)$  for CT; the shaded region where $F(k) \approx D(k)$ is the inertial range.}
\label{fig:force_spectrum_CT}
\end{figure}

 In Fig.~\ref{fig:spectra_flux_RB}{\color{blue}(a)} we plot the  KE and entropy spectra that indicates  Kolmogorov (KO) scaling, i.e. $E_{u}(k)  \sim k^{-5/3}$ and $E_{\theta}(k)  \sim k^{-5/3}$, for convective turbulence.  Our spectrum results are consistent with the KE and entropy fluxes computations, which are plotted in Fig.~\ref{fig:spectra_flux_RB}{\color{blue}(b)}. The KE flux $\Pi_{u}(k)$ and entropy flux $\Pi_{\theta}(k)$ are constant in the inertial range, $20 < k < 1000$.  We also compute energy supply rate $F(k) = \mathrm{Re}( \langle u_k \theta^*_k\rangle)$ and plot it in Figs.~\ref{fig:force_spectrum} and \ref{fig:force_spectrum_CT}.  We observe that $F(k) > 0$ indicating a positive energy transfer from buoyancy to the kinetic energy.  In Fig.~\ref{fig:force_spectrum_CT}, we  also plot the dissipation rate $D(k)$ and $F(k) -D(k)$. In inertial range, $F(k)$ and $ D(k)$ cancel each other approximately, and hence yield a constant KE flux $\Pi_u(k)$.  Thus, we show that in convective turbulence, the KE exhibits Kolmogorov's spectrum, not BO spectrum, as envisaged in some of the earlier work~\cite{Procaccia:PRL1989,Lvov:PRL1991,Lvov:PD1992,Rubinstein:NASA1994}.
 
\section {Discussions and Conclusions}
\label{sec:conclusions}
It is important to contrast our shell model with earlier ones.  Ours is the first shell model for stably stratified turbulence, and it yields results consistent with the BO scaling predicted by Bolgiano~\cite{Bolgiano:JGR1959} and Obukhov~\cite{Obukhov:DANS1959}.   We also observe an approximate dual  spectrum ($k^{-11/5}$ and $k^{-5/3}$) for the kinetic energy for a limited set of parameters.  

By switching the sign of the density gradient, our shell model transforms from stably stratified flows to convective turbulence.  For convective turbulence, our shell model exhibits Kolmogorov's spectrum in the intermediate range of wavenumbers.   Earlier, Brandenburg~\cite{Brandenburg:PRL1992} and Mingshun and Shida~\cite{Mingshun:PRE1997}  had constructed shell models for  RBC.    Brandenburg's~\cite{Brandenburg:PRL1992} shell model is quite different from ours; he added several new terms in the GOY shell model that leads to both forward and inverse KE fluxes. He observes $E_u(k) \sim k^{-5/3}$  for the forward cascade regime, consistent with our model.  However, an inverse cascade of KE flux yields $E_u(k) \sim k^{-11/5}$, which is consistent with the flux arguments of Kumar {\em et al.}~\cite{Kumar:PRE2014}.  When $\Pi_u(k) < 0$ and $F(k) > 0$, Eq.~(26) of Kumar {\em et al.}~\cite{Kumar:PRE2014} would yield $|\Pi_u(k+\Delta k)| < |\Pi_u(k)|$ that  could possibly yield $|\Pi_u(k)| \sim k^{-4/5}$ (BO scaling).  These arguments need a clearcut validation from numerical simulations.  Ching and Cheng~\cite{Ching:PRE2008} used Brandenburg's shell model and studied multiscaling exponents.  

The shell model ``2" of Mingshun and Shida~\cite{Mingshun:PRE1997} is applicable to neutral stratification, and it is a subset of our shell model.  Mingshun and Shida~\cite{Mingshun:PRE1997} reported Kolmogorov's spectrum for the model 2, hence our model is consistent with the shell model of   Mingshun and Shida~\cite{Mingshun:PRE1997}.

We also remark that the shell models are applicable to three-dimensional isotropic turbulence. The numerical work of Kumar {\em et al.}~\cite{Kumar:PRE2014} focusses on Froude number of the order of unity that yields somewhat isotropic flow.  This is the reason why our shell model is consistent with the numerical results of Kumar {\em et al.}~\cite{Kumar:PRE2014}. However, our present shell model   is not expected to work for anisotropic stably stratified flows studied earlier for which the Froude number is quite low~\cite{Lindborg:JFM2006,Vallgren:PRL2011}.  A modification of our shell model to two-dimensional flows may work for  the aforementioned quasi two-dimensional systems. 

In summary,  we constructed a unified shell model for the buoyancy driven turbulence that yields BO scaling for stably stratified flows, but Kolmogorov's spectrum for convective turbulence.   Such low dimensional models have strong utility since they can be used to explore highly non-linear regimes which are inaccessible to numerical simulations and experiments.  

\begin{acknowledgments}
We thank Sagar Chakraborty for valuable suggestions, and Pankaj Mishra for performing initial set of numerical simulations.   Our numerical simulations were performed at {\em HPC2013} and {\em Chaos} clusters of IIT Kanpur. This work was supported by a research grant (Grant No. SERB/F/3279) from Science and Engineering Research Board, India. 
\end{acknowledgments}


\begin{thebibliography}{20}
\expandafter\ifx\csname natexlab\endcsname\relax\def\natexlab#1{#1}\fi
\expandafter\ifx\csname bibnamefont\endcsname\relax
  \def\bibnamefont#1{#1}\fi
\expandafter\ifx\csname bibfnamefont\endcsname\relax
  \def\bibfnamefont#1{#1}\fi
\expandafter\ifx\csname citenamefont\endcsname\relax
  \def\citenamefont#1{#1}\fi
\expandafter\ifx\csname url\endcsname\relax
  \def\url#1{\texttt{#1}}\fi
\expandafter\ifx\csname urlprefix\endcsname\relax\def\urlprefix{URL }\fi
\providecommand{\bibinfo}[2]{#2}
\providecommand{\eprint}[2][]{\url{#2}}


\bibitem[{\citenamefont{{Frisch}}(2011)}]{Frisch:book}
\bibinfo{author}{\bibfnamefont{U.}~\bibnamefont{{Frisch}}},
  \emph{\bibinfo{title}{Turbulence: The Legacy of A N Kolmogorov}}
  (\bibinfo{publisher}{Cambridge University Press},
  \bibinfo{address}{Cambridge}, \bibinfo{year}{2011}).

\bibitem[{\citenamefont{{Ditlevsen}}(2011)}]{Ditlevsen:Book}
\bibinfo{author}{\bibfnamefont{P.}~\bibnamefont{{Ditlevsen}}},
  \emph{\bibinfo{title}{Turbulence and Shell Models}}
  (\bibinfo{publisher}{Cambridge University Press},
  \bibinfo{address}{Cambridge}, \bibinfo{year}{2011}).

\bibitem[{\citenamefont{Biferale}(2003)}]{Biferale:ARFM2003}
\bibinfo{author}{\bibfnamefont{L.}~\bibnamefont{Biferale}},
  \bibinfo{journal}{Ann. Rev. of Fluid Mech.}
  \textbf{\bibinfo{volume}{35}}, \bibinfo{pages}{441} (\bibinfo{year}{2003}).

\bibitem[{\citenamefont{{Brandenburg}}(1992)}]{Brandenburg:PRL1992}
\bibinfo{author}{\bibfnamefont{A.}~\bibnamefont{{Brandenburg}}},
  \bibinfo{journal}{Phys. Rev. Lett.} \textbf{\bibinfo{volume}{69}},
  \bibinfo{pages}{605} (\bibinfo{year}{1992}).

\bibitem[{\citenamefont{{Mingshun} and {Shida}}(1997)}]{Mingshun:PRE1997}
\bibinfo{author}{\bibfnamefont{J.}~\bibnamefont{{Mingshun}}} \bibnamefont{and}
  \bibinfo{author}{\bibfnamefont{L.}~\bibnamefont{{Shida}}},
  \bibinfo{journal}{Phys. Rev. E} \textbf{\bibinfo{volume}{56}},
  \bibinfo{pages}{441} (\bibinfo{year}{1997}).

\bibitem[{\citenamefont{{Ching} and {Cheng}}(2008)}]{Ching:PRE2008}
\bibinfo{author}{\bibfnamefont{E.~S.~C.} \bibnamefont{{Ching}}}
  \bibnamefont{and} \bibinfo{author}{\bibfnamefont{W.~C.}
  \bibnamefont{{Cheng}}}, \bibinfo{journal}{Phys. Rev. E.}
  \textbf{\bibinfo{volume}{77}}, \bibinfo{pages}{015303}
  (\bibinfo{year}{2008}).

\bibitem[{\citenamefont{Kumar et~al.}(2014)\citenamefont{Kumar, Chatterjee, and
  Verma}}]{Kumar:PRE2014}
\bibinfo{author}{\bibfnamefont{A.}~\bibnamefont{Kumar}},
  \bibinfo{author}{\bibfnamefont{A.~G.} \bibnamefont{Chatterjee}},
  \bibnamefont{and} \bibinfo{author}{\bibfnamefont{M.~K.} \bibnamefont{Verma}},
  \bibinfo{journal}{Phys. Rev. E} \textbf{\bibinfo{volume}{90}},
  \bibinfo{pages}{023016} (\bibinfo{year}{2014}).
  
  
\bibitem[{\citenamefont{{Bolgiano}}(1959)}]{Bolgiano:JGR1959}
\bibinfo{author}{\bibfnamefont{R.}~\bibnamefont{{Bolgiano}}},
  \bibinfo{journal}{J. Geophys. Res.} \textbf{\bibinfo{volume}{64}},
  \bibinfo{pages}{2226} (\bibinfo{year}{1959}).

\bibitem[{\citenamefont{{Obukhov}}(1959)}]{Obukhov:DANS1959}
\bibinfo{author}{\bibfnamefont{A.~N.} \bibnamefont{{Obukhov}}},
  \bibinfo{journal}{Dokl. Akad. Nauk SSSR} \textbf{\bibinfo{volume}{125}},
  \bibinfo{pages}{1246} (\bibinfo{year}{1959}).

\bibitem[{\citenamefont{{Lohse} and {Xia}}(2010)}]{Lohse:ARFM2010}
\bibinfo{author}{\bibfnamefont{D.}~\bibnamefont{{Lohse}}} \bibnamefont{and}
  \bibinfo{author}{\bibfnamefont{K.~Q.} \bibnamefont{{Xia}}},
  \bibinfo{journal}{Ann. Rev. Fluid Mech.} \textbf{\bibinfo{volume}{42}},
  \bibinfo{pages}{335} (\bibinfo{year}{2010}).


\bibitem[{\citenamefont{{Procaccia} and {Zeitak}}(1989)}]{Procaccia:PRL1989}
\bibinfo{author}{\bibfnamefont{I.}~\bibnamefont{{Procaccia}}} \bibnamefont{and}
  \bibinfo{author}{\bibfnamefont{R.}~\bibnamefont{{Zeitak}}},
  \bibinfo{journal}{Phys. Rev. Lett.} \textbf{\bibinfo{volume}{62}},
  \bibinfo{pages}{2128} (\bibinfo{year}{1989}).

\bibitem[{\citenamefont{{L'vov}}(1991)}]{Lvov:PRL1991}
\bibinfo{author}{\bibfnamefont{V.~S.} \bibnamefont{{L'vov}}},
  \bibinfo{journal}{Phys. Rev. Lett.} \textbf{\bibinfo{volume}{67}},
  \bibinfo{pages}{687} (\bibinfo{year}{1991}).

\bibitem[{\citenamefont{{L'vov} and {Falkovich}}(1992)}]{Lvov:PD1992}
\bibinfo{author}{\bibfnamefont{V.~S.} \bibnamefont{{L'vov}}} \bibnamefont{and}
  \bibinfo{author}{\bibfnamefont{G.~E.} \bibnamefont{{Falkovich}}},
  \bibinfo{journal}{Physica D} \textbf{\bibinfo{volume}{57}},
  \bibinfo{pages}{85} (\bibinfo{year}{1992}).

\bibitem[{\citenamefont{{Rubinstein}}(1994)}]{Rubinstein:NASA1994}
\bibinfo{author}{\bibfnamefont{R.}~\bibnamefont{{Rubinstein}}},
  \bibinfo{journal}{NASA Technical Memorandum 1066602}  (\bibinfo{year}{1994}).
  


\bibitem[{\citenamefont{{Niemela} et~al.}(2000)\citenamefont{{Niemela},
  {Skrbek}, {Sreenivasan}, and {Donnelly}}}]{Niemela:NATURE2000}
\bibinfo{author}{\bibfnamefont{J.~J.} \bibnamefont{{Niemela}}},
  \bibinfo{author}{\bibfnamefont{L.}~\bibnamefont{{Skrbek}}},
  \bibinfo{author}{\bibfnamefont{K.~R.} \bibnamefont{{Sreenivasan}}},
  \bibnamefont{and} \bibinfo{author}{\bibfnamefont{R.~J.}
  \bibnamefont{{Donnelly}}}, \bibinfo{journal}{Nature}
  \textbf{\bibinfo{volume}{404}}, \bibinfo{pages}{837} (\bibinfo{year}{2000}).
  
\bibitem[{\citenamefont{{Zhang} et~al.}(2005)\citenamefont{{Zhang}, {Wu}, and
  {Xia}}}]{Zhang:PRL2005}
\bibinfo{author}{\bibfnamefont{J.}~\bibnamefont{{Zhang}}},
  \bibinfo{author}{\bibfnamefont{X.~L.} \bibnamefont{{Wu}}}, \bibnamefont{and}
  \bibinfo{author}{\bibfnamefont{K.~Q.} \bibnamefont{{Xia}}},
  \bibinfo{journal}{Phys. Rev. Lett.} \textbf{\bibinfo{volume}{94}},
  \bibinfo{pages}{174503} (\bibinfo{year}{2005}).


\bibitem[{\citenamefont{{Mishra} and {Verma}}(2010)}]{Mishra:PRE2010}
\bibinfo{author}{\bibfnamefont{P.~K.} \bibnamefont{{Mishra}}} \bibnamefont{and}
  \bibinfo{author}{\bibfnamefont{M.~K.} \bibnamefont{{Verma}}},
  \bibinfo{journal}{Phys. Rev. E} \textbf{\bibinfo{volume}{81}},
  \bibinfo{pages}{056316} (\bibinfo{year}{2010}).
  
  
\bibitem[{\citenamefont{{Borue} and {Orszag}}(1997)}]{Borue:JSC1997}
\bibinfo{author}{\bibfnamefont{V.}~\bibnamefont{{Borue}}} \bibnamefont{and}
  \bibinfo{author}{\bibfnamefont{S.~A.} \bibnamefont{{Orszag}}},
  \bibinfo{journal}{J. Sci. Comput.} \textbf{\bibinfo{volume}{12}},
  \bibinfo{pages}{305} (\bibinfo{year}{1997}).

\bibitem[{\citenamefont{{L'vov} et~al.}(1998)\citenamefont{{L'vov},
  {Podivilov}, {Pomyalov}, {Procaccia}, and {Vandembroucq}}}]{Lvov:PRE1998}
\bibinfo{author}{\bibfnamefont{V.}~\bibnamefont{{L'vov}}},
  \bibinfo{author}{\bibfnamefont{E.}~\bibnamefont{{Podivilov}}},
  \bibinfo{author}{\bibfnamefont{A.}~\bibnamefont{{Pomyalov}}},
  \bibinfo{author}{\bibfnamefont{I.}~\bibnamefont{{Procaccia}}},
  \bibnamefont{and}
  \bibinfo{author}{\bibfnamefont{D.}~\bibnamefont{{Vandembroucq}}},
  \bibinfo{journal}{Phys. Rev. E} \textbf{\bibinfo{volume}{58}},
  \bibinfo{pages}{1811} (\bibinfo{year}{1998}).


\bibitem[{\citenamefont{{Gledzer}}(1973)}]{Gledzer:SPD1973}
\bibinfo{author}{\bibfnamefont{E.}~\bibnamefont{{Gledzer}}},
  \bibinfo{journal}{Sov. Phys. Dokl.} \textbf{\bibinfo{volume}{18}},
  \bibinfo{pages}{216} (\bibinfo{year}{1973});
  \bibinfo{author}{\bibfnamefont{K.}~\bibnamefont{{Ohkitani}}} \bibnamefont{and}
  \bibinfo{author}{\bibfnamefont{M.}~\bibnamefont{{Yamada}}},
  \bibinfo{journal}{Prog. Theor. Phys.} \textbf{\bibinfo{volume}{81}},
  \bibinfo{pages}{329} (\bibinfo{year}{1989}).

\bibitem[{\citenamefont{{Verma}}(2004)}]{Verma:PR2004}
\bibinfo{author}{\bibfnamefont{M.~K.} \bibnamefont{{Verma}}},
 \bibinfo{journal}{Phys. Rep.} \textbf{\bibinfo{volume}{401}},
  \bibinfo{pages}{229} (\bibinfo{year}{2004}).


\bibitem[{\citenamefont{Sankar~Ray et~al.}(2008)\citenamefont{Sankar~Ray,
  Mitra, and Pandit}}]{SankarRay:2008bc}
\bibinfo{author}{\bibfnamefont{S.}~\bibnamefont{S.~Ray}},
  \bibinfo{author}{\bibfnamefont{D.}~\bibnamefont{Mitra}}, \bibnamefont{and}
  \bibinfo{author}{\bibfnamefont{R.}~\bibnamefont{Pandit}},
  \bibinfo{journal}{New Journal of Physics} \textbf{\bibinfo{volume}{10}},
  \bibinfo{pages}{033003} (\bibinfo{year}{2008}).

\bibitem[{\citenamefont{Chakraborty et~al.}(2010)\citenamefont{Chakraborty,
  Jensen, and Sarkar}}]{Chakraborty:2010ij}
\bibinfo{author}{\bibfnamefont{S.}~\bibnamefont{Chakraborty}},
  \bibinfo{author}{\bibfnamefont{M.~H.} \bibnamefont{Jensen}},
  \bibnamefont{and} \bibinfo{author}{\bibfnamefont{A.}~\bibnamefont{Sarkar}},
  \bibinfo{journal}{The European Physical Journal B - Condensed Matter}
  \textbf{\bibinfo{volume}{73}}, \bibinfo{pages}{447} (\bibinfo{year}{2010}).


  
  
\bibitem[{\citenamefont{{Lindborg}}(2006)}]{Lindborg:JFM2006}
\bibinfo{author}{\bibfnamefont{E.}~\bibnamefont{{Lindborg}}},
  \bibinfo{journal}{J. Fluid Mech.} \textbf{\bibinfo{volume}{550}},
  \bibinfo{pages}{207} (\bibinfo{year}{2006}).

\bibitem[{\citenamefont{{Vallgren} et~al.}(2011)\citenamefont{{Vallgren},
  {Deusebio}, and {Lindborg}}}]{Vallgren:PRL2011}
\bibinfo{author}{\bibfnamefont{A.}~\bibnamefont{{Vallgren}}},
  \bibinfo{author}{\bibfnamefont{E.}~\bibnamefont{{Deusebio}}},
  \bibnamefont{and}
  \bibinfo{author}{\bibfnamefont{E.}~\bibnamefont{{Lindborg}}},
  \bibinfo{journal}{Phys. Rev. Lett.} \textbf{\bibinfo{volume}{107}},
  \bibinfo{pages}{268501} (\bibinfo{year}{2011}).






\end{thebibliography}
\end{document}